\documentclass[manuscript]{acmart}
\AtBeginDocument{%
  \providecommand\BibTeX{{%
    \normalfont B\kern-0.5em{\scshape i\kern-0.25em b}\kern-0.8em\TeX}}}

\usepackage{multirow}
\usepackage{balance}

\newtheorem{prop}{Proposition}
\copyrightyear{2022} 
\acmYear{2022} 
\setcopyright{acmcopyright}\acmConference[RecSys '22]{Sixteenth ACM Conference on Recommender Systems}{September 18--23, 2022}{Seattle, WA, USA}
\acmBooktitle{Sixteenth ACM Conference on Recommender Systems (RecSys '22), September 18--23, 2022, Seattle, WA, USA}
\acmPrice{15.00}
\acmDOI{10.1145/3523227.3546760}
\acmISBN{978-1-4503-9278-5/22/09}

\begin{document}
\fancyhead{}

\title[TinyKG]{TinyKG: Memory-Efficient Training Framework for  Knowledge Graph Neural Recommender Systems}

\author{Huiyuan Chen}
\email{hchen@visa.com}
\affiliation{%
  \institution{Visa Research}
  \country{USA}
}

\author{Xiaoting Li}
\email{xiaotili@visa.com}
\affiliation{%
  \institution{Visa Research}
  \country{USA}
}

\author{Kaixiong Zhou}
\email{Kaixiong.Zhou@rice.edu}
\affiliation{%
  \institution{Rice University}
  \country{USA}
}

\author{Xia Hu}
\email{xia.hu@rice.edu}
\affiliation{%
  \institution{Rice University}
  \country{USA}
}

\author{Chin-Chia Michael Yeh}
\email{miyeh@visa.com}
\affiliation{%
  \institution{Visa Research}
  \country{USA}
}

\author{Yan Zheng}
\email{yazheng@visa.com}
\affiliation{%
  \institution{Visa Research}
  \country{USA}
}

\author{Hao Yang}
\email{haoyang@visa.com}
\affiliation{%
  \institution{Visa Research}
  \country{USA}
}


\begin{abstract}
There has been an explosion of interest in designing various Knowledge Graph Neural Networks (KGNNs), which achieve state-of-the-art performance and provide great explainability for recommendation. The promising performance is mainly resulting from their capability of capturing high-order proximity messages over the knowledge graphs. However, training KGNNs at scale is challenging due to the high memory usage. In the forward pass, the automatic differentiation engines (\textsl{e.g.}, TensorFlow/PyTorch) generally need to cache all intermediate activation maps in order to compute gradients in the backward pass, which leads to a large GPU memory footprint. Existing work solves this problem by utilizing multi-GPU distributed frameworks. Nonetheless, this poses a practical challenge when seeking to deploy KGNNs in memory-constrained environments, especially for industry-scale  graphs.
 
Here we present TinyKG, a memory-efficient GPU-based training framework for KGNNs for the tasks of recommendation. Specifically, TinyKG uses exact activations in the forward pass while storing a quantized version of activations in the GPU buffers. During the backward pass, these low-precision activations are dequantized back to full-precision tensors, in order to compute gradients. To reduce the quantization errors, TinyKG applies a simple yet effective quantization algorithm to compress the activations, which ensures unbiasedness with low variance. As such, the training memory footprint of KGNNs is largely reduced with negligible accuracy loss. To evaluate the performance of our TinyKG, we conduct comprehensive experiments on  real-world datasets. We found that our TinyKG with INT2 quantization aggressively reduces the memory footprint of activation maps with $7 \times$, only with $2\%$ loss in accuracy, allowing us to deploy KGNNs on memory-constrained devices.

\end{abstract}



\maketitle

\section{Introduction}

Knowledge-aware recommender systems, leveraging the power of Knowledge Graphs (KGs)~\cite{bordes2013translating,yang2014embedding}, have recently gained great attention as they achieve state-of-the-art performance in graph-based recommendation~\cite{wang2019knowledge,wang2019kgat,sun2018recurrent,zhang2016collaborative,wang2020ckan}.  A core benefit of KGs is that they are able to provide high-order connectivity information among items via different types of relations. Such multi-type relations can be seamlessly integrated with user-item interactions, which largely alleviates the data sparsity issues in traditional recommender systems~\cite{zhang2016collaborative,chen2021tops,chen2021structured,wang2019knowledge,wang2019kgat,cao2019unifying}. For instance, CKE~\cite{zhang2016collaborative} combines collaborative filtering with structural knowledge, textual information, and visual signals in a unified framework. Jointly learning the multi-modal heterogeneous graph significantly boosts the quality of recommender systems.

Recently, Knowledge Graph Neural Networks (KGNNs) become one of the most popular graph-based models in recommendation~\cite{cao2019unifying,chen2022graph,wang2022improving,wang2019knowledge,wang2019kgat,wang2020ckan,wang2021learning}. KGNNs use a message-passing mechanism over the KGs, which can be summarized into three stages: i) The input KG is encoded into an embedding space where each KG entity (\textsl{e.g.}, users, items, and attributes) is represented by a low-dimensional vector; ii) Each layer updates the representation of each entity by recursively aggregating and transforming over the representations of its neighbors in the KG; iii) A readout layer is used to obtain the final representation of each entity for the downstream tasks (\textsl{e.g.}, link prediction). For example, KGNN-LS~\cite{wang2019knowledge} transforms the KGs into a user-specific weighted graph and then adopts graph convolution to compute personalized item embeddings with label smoothness regularization. KGAT~\cite{wang2019kgat} recursively propagates the embeddings from a node’s neighbors to refine the  node’s embedding and employs an attention mechanism to discriminate the importance of the neighbors. The success of KGNNs shows that capturing high-order proximity messages over multi-hop neighbors in the KGs is essential for the tasks of recommendation.

Despite their promising performance, training KGNNs at scale is still a challenging problem due to its high computational costs and memory footprint. Modern parallel processors (\textsl{e.g.}, GPUs) often have limited high bandwidth memory capacity. One straightforward strategy is to reduce the batch size to fit the capacity. Nevertheless, a small training batch size leads to poor compute saturation and may cause side effects for both convergence and accuracy~\cite{goyal2017accurate,l.2018dont}. 

Alternatively, various distributed training frameworks \cite{zhu2019graphvite,zheng2020dgl,lerer2019pytorch} have been proposed to parallelize the computations across multiple CPU/GPU accelerators for large-scale KGs. For example, DGL-KE~\cite{zheng2020dgl} and PBG~\cite{lerer2019pytorch} are two popular distributed frameworks for traditional KG embedding models such as TransE~\cite{bordes2013translating} and DistMult~\cite{yang2014embedding}. These methods treat each \textsl{head-relation-tail} triplet in KGs independently so that the input KGs can be partitioned easily and models can be trained in parallel. However, such single-hop distributed frameworks cannot be trivially used for training multi-hop GNN-based KG models. KGNNs require traversing multiple relations in KGs, which span different partitions to learn more complex dependencies among entities~\cite{ren2021smore}. Moreover, distributed KG systems often require high latency, prohibiting their deployments on resource-limited devices.

\vspace{5pt}
\noindent\textbf{Present Work.} The extensive memory of KGNNs stems from the fact that all activations (\textsl{a.k.a.}, feature maps) in the forward pass need to be stored for gradient computation in the back propagation. Thus, training an $L$-layer KGNNs requires to cache all $L$ layers' intermediate activations, which dominates the GPU memory.

Inspired by recent activation compressed training techniques~\cite{chakrabarti2019backprop,fu2020don,chen2021actnn,NEURIPS2020_81f7acab,liu2022gact}, we present TinyKG, a memory-efficient GPU-based training framework for KGNNs for the tasks of recommendation. In particular, TinyKG uses full-precision activations (\textsl{e.g.}, 32-bit floating point (FP32)) during the forward pass while storing a quantized version of activations (\textsl{e.g.}, 2-bit integer (INT2)) in the GPU buffers. During the backward pass, these low-precision activations are dequantized back to full-precision tensors to compute gradients. As such, our TinyKG largely reduces the memory footprint during training, allowing a larger batch size  to fully utilize the power of neural message-passing mechanisms.
 
However, the quantization procedure introduces additional bias and variance, \textsl{i.e.}, the gap between the full-precision values and their quantized values, which inevitably affect the convergence and accuracy of KGNNs. To reduce the quantization error, TinyKG applies a uniform quantization with a stochastic rounding algorithm to compress the activations. We further show that our quantized strategy ensures unbiasedness with low variance. Therefore, the performance of quantized KGNNs is comparable to their original backbones without huge accuracy loss. We conduct extensive experiments to evaluate the effectiveness of the proposed TinyKG. 

Our major contributions are summarized as follows:
 \begin{itemize}
    \item We propose a memory-saving training framework for KGNNs, namely TinyKG, which supports low-bit activation maps in the backward pass. As such, TinyKG  can be easily adapted to many KGNN-based projects.
    \item We introduce a quantization strategy to efficiently compress the activations for back propagation. Besides, we show that our quantization algorithm is unbiased with well bounded variance, which performs well with small time overhead.
    \item We systematically analyze the TinyKG in terms of memory saving, time overhead, and accuracy loss on three public datasets. The results demonstrate that TinyKG can largely reduce memory footprint during training, with negligible loss in accuracy. Generally, TinyKG with INT2 quantization reduces the memory footprint of activation maps with $7 \times$, and  only with $2\%$ loss in accuracy.
\end{itemize}

\section{Related Work}
Our work is related to two lines of research: {Knowledge-aware Recommendation} and {Scalable Graph Training}.  In this section, we briefly go through several prior efforts and discuss their limitations. 

\subsection{Knowledge-aware Recommendation}

Knowledge-aware recommender systems have been successfully utilized to provide complementary information to alleviate the data sparsity or cold start issues~\cite{zhang2016collaborative,wang2019knowledge,wang2019kgat}. The core idea is to transform the entities and relations of KGs into a compact embedding space, where one can impute the missing links/relations in the KGs, such as user-item links. Compared with triple-based models~\cite{zhang2016collaborative,bordes2013translating,chen2020learning,yang2014embedding,cao2019unifying}, such as TransE~\cite{bordes2013translating} and DistMult~\cite{yang2014embedding}, KGNNs provide a class of more powerful architectures that are efficient for capturing multi-hop dependencies over the entire KGs ~\cite{cao2019unifying,wang2019knowledge,wang2019kgat,wang2020ckan,wang2021learning}. Basically, KGNNs follow a recursive aggregation mechanism where each entity aggregates information from its immediate neighbors repeatedly, resulting in better performance.

There are several popular KGNNs in recommender systems,  including R-GCN~\cite{schlichtkrull2018modeling}, KGCN~\cite{wang2019knowledge}, KGAT~\cite{wang2019kgat}, CKAN~\cite{wang2020ckan}, and KGIN~\cite{wang2021learning}. For example,  R-GCN~\cite{schlichtkrull2018modeling} is the first study to show that the GNN framework can be applied for modeling multi-relational data in the KGs.  CKAN~\cite{wang2020ckan} employs a heterogeneous propagation strategy to encode both collaborative user-item signals and knowledge-aware signals, and utilizes attention mechanisms to discriminate the contributions of different neighbors. Recently, KGIN~\cite{wang2021learning} considers both user-item relationships at finer granularity of intents and long-range semantics of relational paths under the message-passing paradigm, which further improves the accuracy and explainability for recommendation.

Although significant research has focused on developing different  architectures of KGNNs, reducing the training memory footprint of KGNNs is much less studied. As the message-passing schema in KGNNs needs to propagate entities' embeddings using multi-hop neighbors, the training of KGNNs is often slow and requires lots of memory. This poses a challenge when seeking to deploy KGNNs for industry-scale graphs~\cite{ren2021smore,teru2020inductive}. 

In this work, we focus on developing a general memory-saving training framework for KGNNs, by observing that most of the memory usage is mainly dominated by the activation maps during the back propagation. Therefore, we propose a new implementation for back propagation by compressing the activation maps, which significantly reduces the GPU memory usage. Interestingly, our memory-saving training framework can be seamlessly applied to many KGNNs since our TinyKG does not need to change the  architectures of existing KGNNs.

\subsection{Scalable Graph Training}

The size of modern KGs has grown exponentially, \textsl{i.e.}, Microsoft Academic Graph has over $8$ billion triples containing information about scientific publications and entities of related entity types~\cite{farber2019microsoft}. Scaling up KG models is thus a critical need for massive KGs in the industry. There are many distributed frameworks for single-hop KG completion, where triples can be partitioned easily~\cite{zhu2019graphvite,zheng2020dgl,lerer2019pytorch}. However, these frameworks cannot be directly used for training KGNNs due to complex multi-hop dependencies in the message-passing schema, especially when KGNNs go deep~\cite{ramezani2020gcn, fey2021gnnautoscale}. Recently, SMORE~\cite{ren2021smore}, the first distributed framework for multi-hop KG models, is built upon a shared memory environment with multiple GPUs, while storing embedding parameters in the CPU memory to overcome the limited GPU memory. Our proposed TinyKG is orthogonal to SMORE and can be used to additionally reduce the activation maps of SMORE in the back propagation. We leave the potential combination as our future study.

On the other hand, great efforts have been made to train deep neural networks in resource-constraint scenarios~\cite{han2015deep_compression,mostafa2019parameter,liu2021exact,chen2016training,micikevicius2018mixed,fu2020don,chen2021actnn,yeh2021embedding}. For example, model compression~\cite{han2015deep_compression} and gradient compression~\cite{lin2018dgc} are able to reduce the storage and communication overhead by compressing the weights and gradients. Gradient checkpointing~\cite{chen2016training} trades computation for memory by dropping some of activations in the forward pass, and recomputing them in the backward pass.  Swapping~\cite{meng2017training,huang2020swapadvisor} fully utilizes a large amount of host CPU memory by exchanging tensors between CPU and GPU. Nevertheless, swapping techniques do not fit memory-constrained devices as there is no sufficient host memory. Quantization-aware Training~\cite{tailor2021degreequant,jin2022fnet,ding2021vq} generally aims to improve the model efficiency at the training or inference stage. However, many existing frameworks use full-precision data type to simulate the effect of real quantization (\textsl{a.k.a.} fake quantization) since many CUDA kernels cannot directly support low-bit operators/tensors. Also, the convergence behavior of quantization-aware training is still not well understood, limiting their generality.

In contrast, Activation Compressed Training~\cite{chakrabarti2019backprop,fu2020don,chen2021actnn,evans2021ac,liu2022gact} becomes a promising technique to reduce the training memory footprint as it only considers \textsl{storage}, allowing more flexible quantization strategies for better compression. This technique has been successfully used to train ResNet with $2$-bit activations, reducing memory footprint by $12\times$, and enabling a $14\times$ larger training batch size in practice~\cite{chen2021actnn,fu2020don}. However, there is no existing work that extends this direction to KGNNs and analyzes its feasibility.

Our TinyKG is built upon this line of work, and provides a memory-efficient GPU implementation to support common graph convolutional operators in the KGNNs.

\section{The Proposed TinyKG}
\subsection{Problem Setup}

We begin by describing the problem of knowledge-aware recommendation and introducing the notations.

\vspace{5pt}
\noindent\textbf{User-Item Graph.} In this work, we focus on learning user preferences from the implicit feedback (\textsl{e.g.}, click, comment, purchase, etc.). To be specific, we have a set of users $\mathcal{U}=\{u\}$ and a set of items $\mathcal{V}=\{v\}$. Let $\mathbf{Y}$ be the user-item interaction matrix, where $y_{uv}=1$ indicates that user $u$ has engaged with item $v$ before, and $y_{uv}=0$ otherwise.

\vspace{5pt}
\noindent\textbf{Knowledge Graph.}
A KG is a multi-relational graph where nodes denote entities, and edges correspond to relations between entities. An edge in a KG represents a fact stored in the form of  (\textsl{subject}, \textsl{predicate}, \textsl{object}). In recommendation scenarios, A KG often contains the side information, such as item attributes, taxonomy, or external knowledge. Formally, a KG is represented as $\mathcal{G} = \{(h, r, t)| h, t \in \mathcal{E}, r \in \mathcal{R}\}$, in which each triple $(h,r,t)$ indicates that a relation $r$ exists from head entity $h$ to tail entity $t$, $\mathcal{E}$ and $\mathcal{R}$ are the set of entities and relations in the KG, respectively. For example, the triple (\textsl{Tom Hanks}, \textsl{ActorOf}, \textsl{Forrest Gump}) states the fact that Tom Hanks is an actor of the movie Forrest Gump.

In knowledge-aware recommendation, the user-item graph  $\mathbf{Y}$ can be  seamlessly integrated with the knowledge graph $\mathcal{G}$ based on the item-entity alignment. An item $v \in \mathcal{V}$ corresponds to an entity $e \in \mathcal{E}$ (\textsl{e.g.}, the item "Forrest Gump" also appears in the KG as an entity). The set of entities $ \mathcal{E}$ consists of items and item attributes. Normally, the KG provides complementary information for user-item graph, which highly alleviates the data sparsity or cold start issues.

\vspace{5pt}
\noindent\textbf{Task Description.}
Given the user-item interaction graph $\mathbf{Y}$ and the knowledge graph $\mathcal{G}$, the task of knowledge-aware recommendation is to predict how likely a user would adopt an item that he/she has not interacted before. That is:
\begin{equation*}
		\hat{y}_{uv} = \text{KG-Model} (u, v |\mathbf{A}^\mathcal{R}, \mathbf{\Theta}),
\end{equation*}
where $\hat{y}_{uv}$ denotes the probability that user $u$ will engage with item $v$, $\mathbf{A}^\mathcal{R}$ represents  the multi-relational matrix that can be constructed from $\mathbf{Y}$ and $\mathcal{G}$, and $\mathbf{\Theta}$ is the model parameters of the KG-Model.

\subsection{Knowledge Graph Neural Networks}
Knowledge Graph Neural Networks (KGNNs) have been shown great potential in improving diversity, accuracy, and explainability for personalized recommendation~\cite{wang2019kgat,wang2019knowledge,wang2021learning}. One of the key benefits of KGNNs is their ability of capturing high-order structural proximity among entities over the KGs, which alleviates the sparsity issue in recommendation.

	\begin{figure*}
		\begin{center}
			\includegraphics[width=0.96\linewidth]{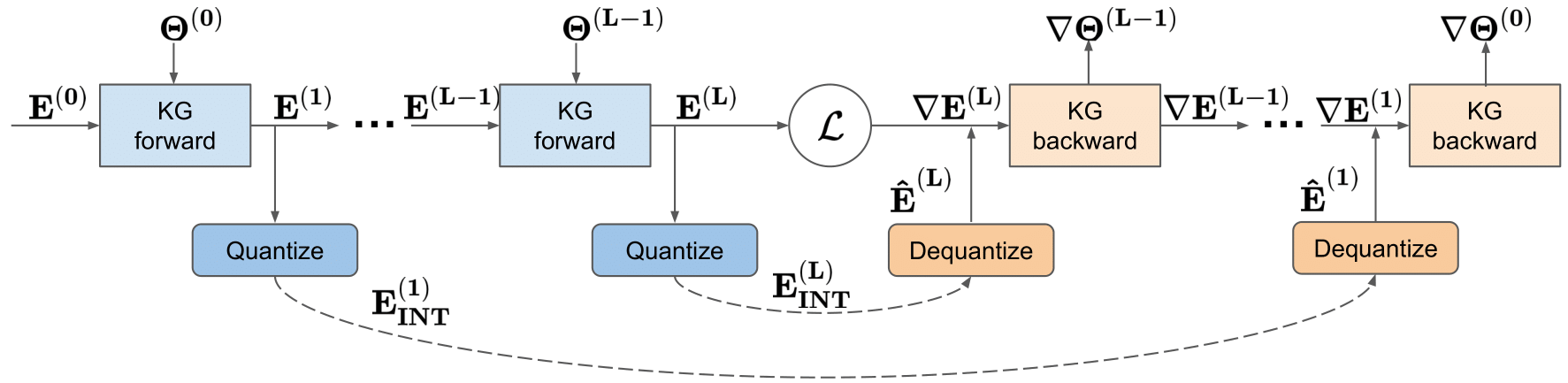}
		\end{center}
		\caption{The pipeline of the proposed TinyKG, where we quantize the full-precision activation maps $\{\mathbf{E}^{(1)}, \cdots, \mathbf{E}^{(L)}\}$  into lower numerical precision values $\{\mathbf{E}_{\text{INT}}^{(1)},\cdots,\mathbf{E}_{\text{INT}}^{(L)}\}$ while still propagating the exact $\{\mathbf{E}^{(1)}, \cdots, \mathbf{E}^{(L)}\}$ during the forward pass. The low-precision activations are then dequantized back to full-precision tensor $\{\hat{\mathbf{E}}^{(1)}, \cdots, \hat{\mathbf{E}}^{(L)}\}$, which are used to compute gradients during the backward pass. Note that TinyKG only caches the quantized activations in the GPU buffers to reduce memory footprint.}
		\label{tinykg}
	\end{figure*}

\noindent\textbf{Message-passing Schema.} Most of KGNNs fit under the message-passing schema, where the representation $\mathbf{e}_v$ of each entity is updated iteratively in each layer by collecting messages from its neighbors in a KG. In particular, the $l$-th layer can be simplified as:
\begin{equation}
		\label{eq:kgcn}
			{\bf E}^{(l+1)} = \text{KG-Layer}  \left( \mathbf{A}^\mathcal{R},	{\bf E}^{(l)},  \mathbf{\Theta}^{(l)} \right), \quad l = 0, 1, \cdots, L-1.
\end{equation}
where ${\bf E}^{(l)} \in \mathbb{R}^{N\times d}$ denotes the $d$-dimensional embeddings of entities at the $l$-th layer, $N$ is the number of entities in the KGs, $L$ is the number of layers, $\mathbf{A}^\mathcal{R}$ denotes the relational matrix that contains multi-type relationships among entities, $\mathbf{\Theta}^{(l)}$ denotes the trainable parameters in the $l$-th layer, and $\text{KG-Layer}(\cdot)$ is the propagation layer, such as graph convolutional layer~\cite{wang2019knowledge}, graph attention layer~\cite{wang2019kgat}, and  path-aware propagation layer~\cite{wang2021learning}. After $L$ layers,  a readout layer may be adopted to generate the final embedding for each entity. Finally, we can use downstream loss with different regularizations to optimize the model parameters~\cite{wang2019kgat,wang2019knowledge,wang2021learning}.

Taking KGNN-LS~\cite{wang2019knowledge} as an example, its aggregation process is: $	{\bf E}^{(l+1)} = \sigma ( \hat{\mathbf{A}}{\bf E}^{(l)} \mathbf{\Theta}^{(l)} )$, where $\hat{\mathbf{A}}$ is the normalized adjacency matrix of $\mathbf{A}^\mathcal{R}$ with self connection, and $\sigma(\cdot)$ is the non-linear function.  Then, its computational graph can be decomposed as:
\begin{equation}
\label{cg}
\begin{aligned}
\textsl{Forward}: \qquad  \mathbf{H}^{(l)}  &= \text{spmm}(\hat{\mathbf{A}}, \mathbf{E}^{(l)})  \; \downarrow \;  \\ \mathbf{J}^{(l)}  &= \text{mm} (\mathbf{H}^{(l)},  \mathbf{\Theta}^{(l)}) \;   \downarrow \;  \\ \mathbf{E}^{(l+1)}  &=  \sigma(\mathbf{J}^{(l)}), \\
\textsl{Backward}: \qquad  \nabla_{\mathbf{J}^{(l)}} &=  \text{ctx}(\mathbf{J}^{(l)}, \nabla_{\mathbf{E}^{(l+1)}})  \; \downarrow\; \\
(\nabla_{\mathbf{H}^{(l)}}, \nabla_{\mathbf{\Theta}^{(l)}}) &=  \text{ctx}(\mathbf{H}^{(l)}, \mathbf{\Theta}^{(l)}, \nabla_{\mathbf{J}^{(l)}} ) \; \downarrow \; \\
 \nabla_{\mathbf{E}^{(l)}} &=  \text{ctx}({\hat{\mathbf{A}}}, \nabla_{\mathbf{H}^{(l)}}),
\end{aligned}
\end{equation}
where $\text{spmm}(\cdot)$ is the sparse-dense matrix multiplication,  $\text{mm}(\cdot)$ is the dense-dense matrix multiplication, $\nabla_{(\cdot)}$ denotes the gradient of activation/parameter that is always taken with respect to the loss $\mathcal{L}$, and $\text{ctx}(\cdot)$ denotes the context  information that needs to be stored in the GPU memory for the backward pass, \textsl{i.e.}, $   \nabla_{\mathbf{\Theta}^{(l)}} =  \text{ctx}(\mathbf{H}^{(l)},  \nabla_{\mathbf{J}^{(l)}} ) ={\mathbf{H}^{(l)}}^\top \nabla_{\mathbf{J}^{(l)}} $. Essentially, the backward pass requires more context information than the forward pass.

\vspace{5pt}
\noindent\textbf{Memory Analysis.} In the inference stage, one can only perform the forward pass of the entire network, the results of the intermediate layers (\textsl{e.g.}, $\mathbf{H}^{(l)}$) can be discarded subsequently. However, in the training stage,  the automatic differentiation engines (\textsl{e.g.,} PyTorch) need to store the following forward-pass variables in the memory to compute the gradients in the backward pass: 
\begin{itemize}
    \item The normalized adjacency matrix $\hat{\mathbf{A}}$:  the matrix $\hat{\mathbf{A}}$  is often very sparse in KGs and only need to be kept once for all $L$ KG-Layers. Thus, the memory footprint of $\hat{\mathbf{A}}$ is trivial in the backward pass with space complexity $\mathcal{O}(|\mathcal{E}|)$, where $|\mathcal{E}|$ denotes the number of edges in the KGs.
    \item The parameter of $\mathbf{\Theta}^{(l)} \in \mathbb{R}^{d\times d}$: the space complexity of $\mathbf{\Theta}^{(l)}$ is independent to the size of KGs (\textsl{e.g.,} $d \ll N$). The model parameters is generally negligible in recommendation  with space complexity $\mathcal{O}(Ld^2)$.
    \item 
All intermediate results $\mathbf{H}^{(l)} \in \mathbb{R}^{N\times d}$, $\mathbf{J}^{(l)} \in \mathbb{R}^{N\times d} $, $\mathbf{E}^{(l)} \in \mathbb{R}^{N\times d}$ (if layer-aggregation is adopted~\cite{wang2019kgat,wang2021learning}): training an $L$-layer KGNN requires to cache all $L$ layers' intermediate outputs with a $\mathcal{O}(LNd)$ space complexity. These intermediate outputs are termed as "activation maps"~\cite{evans2021ac,fu2020don,NEURIPS2020_81f7acab}.
\end{itemize}

From the above analysis, it is the activation maps ($\mathcal{O}(LNd)$) that mainly dominate the GPU memory during training, not the model parameters ($\mathcal{O}(Ld^2)$). Based on this observation, in this work, we aim to reduce the capacity of intermediate activations by using principled quantized techniques to expedite the KGNNs' training.

\subsection{Quantized Activation Maps}

To reduce the memory consumption at the training stage, we present TinyKG, a memory-saving training framework for KGNNs without modifying their original architectures. Instead of saving the full-precision tensors (\textsl{e.g.}, FP32), TinyKG aims to lazily save quantized activation maps with lower numerical precision (\textsl{e.g.}, INT2) in the GPU buffers for back propagation. 

Figure~\ref{tinykg} is an overview of our proposed TinyKG. To be specific, the full-precision activations (\textsl{e.g.}, $\mathbf{E}^{(l)}$ ) is used during the forward pass. Then, the full-precision activations will be compressed into low-precision tensors via quantization
(\textsl{e.g.}, $\mathbf{E}_{\text{INT}}^{(l)}$), which overwrites the exact activations in the GPU buffers. The full-precision activations (\textsl{e.g.}, $\mathbf{E}^{(l)}$ ) can be then discarded in a layer-by-layer order. During the backward pass, TinyKG dequantizes the compressed activations in the GPU buffers back to full-precision tensors (\textsl{e.g.}, $\hat{\mathbf{E}}^{(l)}$). The gradients are then computed based on the dequantized activations. Therefore, the memory required for saving activation maps is highly reduced, enabling training with a larger batch size, or scaling up the model size, on the same GPU device. 

In principle, any compression algorithm, either lossy or lossless, can be used to compress activation maps~\cite{Stock2020And}. In this work, we introduce a simple yet effective quantized strategy to compress activation maps. Besides, we show that our quantized strategy is unbiased with low variance.

\vspace{5pt}
\noindent\textbf{Quantization (Float $\to$ Integer).} Specifically, the activation $\mathbf{E}^{(l)}$ (same for other activations) will be quantized and stored using $b$-bit integers. Let $B = 2^b-1$ be the number of quantization bins, we quantize each  tensor $\mathbf{e}_v^{(l)}$ of $\mathbf{E}^{(l)}$  as:
\begin{equation}
		\label{deq}
{\mathbf{e}^{(l)}_{v_{INT}}}  = \text{Quant}(\mathbf{e}^{(l)}_v)  =\left\lfloor \frac{\mathbf{e}^{(l)}_v - Z^{(l)}_v}{R^{(l)}_v}B \right\rceil,
\end{equation}
where $R^{(l)}_v = \max\{\mathbf{e}_v^{(l)}\} - \min\{\mathbf{e}_v^{(l)}\}$ is the range for $\mathbf{e}_v^{(l)}$, $Z^{(l)}_v = \min\{\mathbf{e}_v^{(l)}\}$ is the offset, ${\mathbf{e}^{(l)}_{v_{INT}}}$ is the compressed activation scaled to $[0, B]$, and $\lfloor \cdot \rceil$ denotes the stochastic rounding operator~\cite{gupta2015deep,courbariaux2015binaryconnect}.  For any scalar $x$, the stochastic rounding can be formulated as:
\begin{equation*}
  \lfloor x\rceil=\left\{
    \begin{array}{ll}
      \lceil x \rceil, & \mbox{with probability $x -\lfloor x \rfloor$},\\
      \lfloor x \rfloor, & \mbox{with probability $1- (x -\lfloor x \rfloor)$},
    \end{array}
  \right.
\end{equation*}
where $\lceil \cdot \rceil$ is the ceil operator and $\lfloor \cdot \rfloor$ is the floor operator.

\vspace{5pt}
\noindent\textbf{Dequantization (Integer $\to$ Float).} During the backward pass,  the compressed activation ${\mathbf{e}^{(l)}_{v_{INT}}}$ is dequantized as:
\begin{equation}
		\label{deq2}
\hat{\mathbf{e}}^{(l)}_v  = \text{Dequant}({\mathbf{e}^{(l)}_{v_{INT}}} )  = \frac{R_v^{(l)}{\mathbf{e}^{(l)}_{v_{INT}}}}{B} + Z_v^{(l)},
\end{equation}
where $\hat{\mathbf{e}}^{(l)}_v$ is a full-precision tensor that are then used to calculate the gradients for back propagation. All operators (\textsl{e.g.}, $\text{spmm}(\cdot)$) are performed in full-precision. Note that the dequantized step is needed since most of the GPUs do not support low-bit operators, rather than full-precision and half-precision operators\footnote{\url{https://pytorch.org/blog/accelerating-training-on-nvidia-gpus-with-pytorch-automatic-mixed-precision/}.}.

\vspace{5pt}
\noindent\textbf{Bias and Variance.} Inspired by recent efforts \cite{connolly2021stochastic,chen2021actnn,liu2021exact,liu2022gact}, we have the following key property of the quantization:
	\begin{prop}
The quantization of Eq. (\ref{deq}) and Eq. (\ref{deq2}) for activation $\mathbf{e}_v^{(l)} \in \mathbb{R}^d$ is unbiased with well bounded variance, and its expectation and variance are:
\begin{equation}
		\label{prop1}
\mathbb{E}[\hat{\mathbf{e}}^{(l)}_v]=\mathbb{E}[\textup{Dequant}(\textup{Quant}(\mathbf{e}^{(l)}_v))] = \mathbf{e}^{(l)}_v, \qquad \textup{Var}[\hat{\mathbf{e}}^{(l)}_v] \le \frac{d[R^{(l)}_v]^2}{4B^2}.
\end{equation}
	\end{prop}

The detailed analysis of Proposition 1 is given in Appendix. As the $\text{Dequant}(\text{Quant}(\cdot))$ is an unbiased quantizer, the computed gradients are also unbiased. Nevertheless, the quantization inevitably imposes additional variance to the gradients during the backward pass, which plays an important role to the convergence behavior. As can be seen, the variance is inversely correlated with number of quantization bins $B$, \textsl{i.e.}, a larger $B$ leads to smaller variance. In the experiments, we will investigate how different values of $B$ affect the model performance (Sec 4.2.3). 

To make our quantized strategy more rigorous, we theoretically show how much extra variance activation compression introduces. Let $\{\nabla_{\mathbf{\Theta}^{(l)}}, \nabla_{\mathbf{E}^{(l)}} \}$
be the full-precision gradients of $\{{\mathbf{\Theta}^{(l)}}, {\mathbf{E}^{(l)}} \}$, and  $\{\hat{\nabla}_{\mathbf{\Theta}^{(l)}}, \hat{\nabla}_{\mathbf{E}^{(l)}} \}$ be the corresponding gradients using the compressed context. Further, we use the notation $\mathbf{G}^{(l \sim m)}_{\mathbf{\Theta}}(\hat{\nabla}_{\mathbf{E}^{(m)}}, \hat{\mathbf{C}}^{(m)})$ to represent the variance introduced by using the compressed context $\hat{\mathbf{C}}^{(m)}$. From Theorem 3 in ~\cite{chen2021actnn}, given an $L$-layer KGNN, we have:
\begin{equation}
\label{eq6}
    \operatorname{Var}\left[\hat{\nabla}_{\mathbf{\Theta}^{(l)}}\right]=\operatorname{Var}\left[\nabla_{\mathbf{\Theta}^{(l)}}\right]+\sum_{m=l}^{L} \mathbb{E}\left[\operatorname{Var}\left[\mathbf{G}_{\mathbf{\Theta}}^{(l \sim m)}\left(\hat{\nabla}_{\mathbf{E}^{(m)}}, \hat{\mathbf{C}}^{(m)}\right) \mid \hat{\nabla}_{\mathbf{E}^{(m)}}\right]\right],
\end{equation}
where $\operatorname{Var}\left[\cdot|\hat{\nabla}_{\mathbf{E}^{(m)}}\right]$ is the conditional variance, and the $\operatorname{Var}\left[\nabla_{\mathbf{\Theta}^{(l)}}\right]$ is the full-precision gradient variance in the original Stochastic Gradient Descent.  Intuitively, the variance introduced by compressed context at different layers will accumulate as KGNNs go deep. Fortunately, most of KGNNs often have shallow architectures (\textsl{e.g.}, $L \le 4$)~\cite{wang2019kgat,wang2019knowledge,wang2021learning}, the gradient variance is thus trivial comparing to deep CNNs in practice~\cite{chen2021actnn}.
 
According to Eq. (\ref{prop1}) and Eq. (\ref{eq6}), we may reduce the numerical precision for free, as the quantization is unbiased and the variance is negligible. This suggests that there is no need to adopt expensive sophisticated quantization strategies, like mixed-precision quantization or non-uniform quantization, as considered in previous work~\cite{chen2021actnn,l.2018dont}. Moreover, it is worth noting that our proposed TinyKG is easily compatible to any KGNNs since it only changes the routine of storage saving in the GPU buffers, rather than their vanilla network architectures.

\section{EXPERIMENTS}

In real-world applications, the deployed models should achieve a balanced trade-off among model performance, speed, and space complexity. In this section, we systematically analyze the proposed TinyKG in terms of memory saving, time overhead, and accuracy loss on three real-world datasets.

\begin{table*}[t]
\caption{The statistics of the benchmark datasets.}
\label{dataset}
\resizebox{0.66\textwidth}{!}
{
\begin{tabular}{cl|r|r|r}
\toprule[1.1pt]
                & \multicolumn{1}{c|}{} & \multicolumn{1}{c|}{Amazon-book} & \multicolumn{1}{c|}{MovieLens-20M} & \multicolumn{1}{c}{Yelp2018} \\ \hline 
                & \#Users               & $70,679$                           & $138,159$                            & $45,919$                       \\
User-Item Graph & \#Items               & $24,915$                           & $16,954$                             & $45,538$                       \\
                & \#Interactions        & $847,733$                          & $13,501,622$                         & $1,185,068$                    \\ \hline 
                & \#Entities            & $88,572$                           & $102,569$                            & $90,961$                       \\
Knowledge Graph & \#Relations           & $39$                               & $32$                                 & $42$                           \\
                & \#Triples             & $2,557,746$                        & $499,474$                            & $1,853,704$                    \\ \toprule[1.1pt]
\end{tabular}}
\end{table*}
\subsection{Experimental Settings}
\subsubsection{\textbf{Datasets}}
We conduct experiments on three publicly available benchmark datasets: Amazon-book, MovieLens-20M, and Yelp, which vary in terms of domain, size, and sparsity:
\begin{itemize}
    \item \textbf{Amazon-book\footnote{\url{http://jmcauley.ucsd.edu/data/amazon}.}:} The dataset contains a large corpus of user reviews, ratings, and product metadata, collected from the Amazon Book category. To guarantee the quality of the dataset, we use the $10$-core setting, retaining users and items with at least ten interactions.
    \item \textbf{MovieLens-20M\footnote{\url{https://grouplens.org/datasets/movielens/20m/}.}:} The dataset is a widely used benchmark dataset in recommendations, which consists of approximately $20$ million explicit ratings (ranging from $1$ to $5$). We transform them into implicit feedback, where each user-item interaction is marked with $1$ if its rating score is greater than $3$, otherwise $0$.
    \item \textbf{Yelp\footnote{\url{https://www.yelp.com/dataset}.}:} This dataset is adopted from the 2018 edition of the Yelp challenge. In this work, we view the local businesses like restaurants and bars as items. Similarly, we use the $10$-core setting to ensure that each user and item have at least ten interactions.
\end{itemize}

In addition to the user-item graph, we directly follow the previous work~\cite{wang2019kgat,wang2019knowledge,wang2020ckan,wang2021learning} to construct an item knowledge graph for each dataset. Then, we merge the user-item graph and the item knowledge graph via item alignments. In particular, we consider triples from the item knowledge graph that are directly related to the entities to align with the items in the user-item graph, regardless of which role it plays (\textsl{i.e.}, subject or object). 

Table~\ref{dataset} briefly summarizes the statistics of those three datasets. For each dataset, we randomly select $80\%$ of interaction history of each user to constitute the training set, and treat the remaining as the test set.
From the training set, we randomly select $10\%$ of interactions as the validation set to tune hyper-parameters~\cite{wang2019kgat,wang2021learning}.


\subsubsection{\textbf{Baselines}}
The main goal of our TinyKG is to reduce the training GPU memory of the KGNNs, not to design new architectures. Therefore, we evaluate our framework on existing state-of-the-art KGNNs, including:
\begin{itemize}
    \item \textbf{KGAT}~\cite{wang2019kgat}: KGAT utilizes an attentive neighborhood aggregation mechanism on the KG to generate user/item representations. As such, it is able to discriminate the important of neighbors during propagation, leading to more accurate and explainable recommendation.

    \item \textbf{KGNN}~\cite{wang2019knowledge}: KGNN first transforms the heterogeneous KG into a user-specific weighted graph, and then computes the personalized item embedding based on graph neural network. It also introduces the label smoothness regularization to  provide better inductive bias.

     \item \textbf{KGIN}~\cite{wang2021learning}: KGIN is the latest state-of-the-art propagation-based recommendation method, which explicitly models user interaction behaviors with latent intents. It adopts a relational path-aware aggregation schema to capture the long-range dependencies in the KG.

\end{itemize}

The baselines are using full-precision training, while the proposed TinyKG will compress their corresponding activation maps in the GPU buffers during training. However, their behaviors are identical at the inference stage.

\subsubsection{\textbf{Evaluation Metrics}} We  follow the previous work~\cite{wang2019kgat,wang2021learning} to conduct the evaluation of top-$K$ recommendation. For each user in the test set, we treat all the items that the user has not interacted with as the negative items.
Then each KGNN model predicts the user's preference scores over all the items, except the positive ones in the training set.
To evaluate the effectiveness of different approaches, we adopt two widely-used top-$K$ evaluation protocols~\cite{wang2019kgat,wang2021learning}: Recall@$K$ and NDCG@$K$. By default, we set $K=20$, and we report the average metrics for all users in the test set.

\subsubsection{\textbf{Implementation Details}} We implement all KGNNs in PyTorch, and run the experiments on a Linux machine with a single NVIDIA Tesla P100 with 16 GB GPU memory. For a fair comparison, we fix the embedding size of the entities (\textsl{e.g.}, Eq. (\ref{eq:kgcn})) as $64$, the training batch size as $1024$, the number of KGNN layer as $3$, the learning rate as $1e^{-3}$, and the optimizer as Adam for all the baselines. For other model-specific hyper-parameters, we use the default settings as suggested in the original papers.

For TinyKG, the common operators  (\textsl{e.g.,} Linear, ReLU, BatchNorm, SPMM, SPMM\_MAX, etc.) are built on the top of existing  frameworks~\cite{chen2021actnn,liu2021exact,pan2021mesa,liu2022gact} with different configurations using CUDA kernels. All the  substitutions within KGNNs can be done automatically with a model converter. Taking ReLU($\cdot$) as an example, we have $\mathbf{y}= \text{ReLU}(\mathbf{x}) = \mathbf{x} \odot \mathbf{1}_{\mathbf{x}>0}$ and $\nabla_\mathbf{y}= \nabla_\mathbf{y} \odot \mathbf{1}_{\mathbf{x}>0}$. Here, ReLU($\cdot$) only needs to store $\mathbf{1}_{\mathbf{x}>0}$ for the backward pass, which takes one bit per element in the buffers. As PyTorch only supports precision down to INT8, our TinyKG can further convert quantized tensors into bit-streams by CUDA kernels to maximize the memory saving. In addition, our TinyKG is able to support both full-precision and half-precision (\textsl{e.g.}, bfloat16) that is thus compatible with Automated Mixed Precision training (AMP)\footnote{\url{https://pytorch.org/docs/stable/amp.html}}, to further reduce the memory consumption.

\begin{table*}[]
\caption{The performance of KGNNs trained on the Amazon-book dataset with compressed activations storing in different precision. All results are averaged over ten random trials. All number are percentage numbers without $\%$.}
\label{table2}
\resizebox{0.8\textwidth}{!}
{\begin{tabular}{crrrrrr}
 \toprule[1.1pt]
Model                  & Metric & FP32(baseline)                       & INT8          & INT4        & INT2        & INT1        \\ \hline
                       & Recall@20 (\%) & $14.85  {\color[HTML]{656565} \pm 0.15}$ & $14.81{\color[HTML]{656565} \pm0.12}$   & $14.78{\color[HTML]{656565} \pm0.14}$ & $14.66{\color[HTML]{656565} \pm0.14}$ & $14.08{\color[HTML]{656565} \pm0.15}$ \\
\multirow{-2}{*}{KGAT} & NDCG@20 (\%)   & $8.12 {\color[HTML]{656565} \pm 0.10}$                         & $8.10 {\color[HTML]{656565} \pm0.12}$   & $8.08{\color[HTML]{656565} \pm0.11}$  & $7.98{\color[HTML]{656565} \pm0.12}$  & $7.65{\color[HTML]{656565} \pm0.17}$  \\ \hline
                       & Recall@20 (\%) & $13.64 {\color[HTML]{656565} \pm 0.16}$                        & $13.61 {\color[HTML]{656565} \pm 0.18}$ & $13.58{\color[HTML]{656565} \pm0.12}$ & $13.46{\color[HTML]{656565} \pm0.12}$ & $12.91{\color[HTML]{656565} \pm0.13}$ \\
\multirow{-2}{*}{KGNN} & NDCG@20 (\%)   & $5.71 {\color[HTML]{656565} \pm 0.09}$                         & $5.70 {\color[HTML]{656565} \pm 0.07}$  & $5.67{\color[HTML]{656565} \pm0.10}$  & $5.62{\color[HTML]{656565} \pm0.09}$  & $5.39{\color[HTML]{656565} \pm0.11}$  \\ \hline
                       & Recall@20 (\%) & $16.89{\color[HTML]{656565} \pm0.16}$                          & $16.85 {\color[HTML]{656565} \pm 0.13}$ & $16.82{\color[HTML]{656565} \pm0.12}$ & $16.65{\color[HTML]{656565} \pm0.12}$ & $15.97{\color[HTML]{656565} \pm0.11}$  \\
\multirow{-2}{*}{KGIN} & NDCG@20 (\%)   & $9.16 {\color[HTML]{656565} \pm 0.12}$                         & $9.14 {\color[HTML]{656565} \pm 0.14}$  & $9.10{\color[HTML]{656565} \pm0.12}$  & $8.99{\color[HTML]{656565} \pm0.15}$  & $8.63{\color[HTML]{656565} \pm0.12}$  \\ \toprule[1.1pt]
\end{tabular}}
\end{table*}

\begin{table*}[]
\caption{The performance of KGNNs trained on the MovieLens-20M dataset with compressed activations storing in different precision. All results are averaged over ten random trials. All number are percentage numbers without $\%$.}
\label{table3}
\resizebox{0.8\textwidth}{!}
{\begin{tabular}{crrrrrr}
 \toprule[1.1pt]
Model                  & Metric & FP32(baseline)                       & INT8          & INT4        & INT2        & INT1        \\ \hline
                       & Recall@20 (\%) & $22.13 {\color[HTML]{656565} \pm 0.13}$ & $22.09{\color[HTML]{656565} \pm 0.11}$  & $21.96{\color[HTML]{656565} \pm0.13}$ & $21.73{\color[HTML]{656565} \pm0.14}$ & $20.85{\color[HTML]{656565} \pm0.17}$ \\
\multirow{-2}{*}{KGAT} & NDCG@20 (\%)   & $15.73 {\color[HTML]{656565} \pm 0.07}$ & $15.70 {\color[HTML]{656565} \pm 0.10}$  & $15.61{\color[HTML]{656565} \pm0.11}$  & $15.45{\color[HTML]{656565} \pm0.12}$  & $14.98{\color[HTML]{656565} \pm0.14}$  \\ \hline
                       & Recall@20 (\%) & $21.04 {\color[HTML]{656565} \pm 0.11}$ & $20.99{\color[HTML]{656565} \pm 0.13}$  & $20.85{\color[HTML]{656565} \pm0.13}$ & $20.63{\color[HTML]{656565} \pm0.11}$ & $20.01{\color[HTML]{656565} \pm0.12}$ \\
\multirow{-2}{*}{KGNN} & NDCG@20 (\%)   & $13.26 {\color[HTML]{656565} \pm 0.10}$ & $13.24 {\color[HTML]{656565} \pm 0.08}$  & $13.15{\color[HTML]{656565} \pm0.10}$  & $13.02{\color[HTML]{656565} \pm0.10}$  & $12.50{\color[HTML]{656565} \pm0.11}$  \\ \hline
                       & Recall@20 (\%) & $24.58 {\color[HTML]{656565} \pm 0.11}$ & $24.54{\color[HTML]{656565} \pm 0.12}$  & $24.39{\color[HTML]{656565} \pm0.13}$ & $24.14{\color[HTML]{656565} \pm0.13}$ & $23.65{\color[HTML]{656565} \pm0.14}$  \\
\multirow{-2}{*}{KGIN} & NDCG@20 (\%)   & $18.05 {\color[HTML]{656565} \pm 0.14}$ & $18.01 {\color[HTML]{656565} \pm 0.15}$  & $17.92{\color[HTML]{656565} \pm0.14}$  & $17.75{\color[HTML]{656565} \pm0.15}$  & $17.21{\color[HTML]{656565} \pm0.12}$  \\ \toprule[1.1pt]
\end{tabular}}
\end{table*}

\begin{table*}[]
\caption{The performance of KGNNs trained on the Yelp2018 dataset with compressed activations storing in different precision. All results are averaged over ten random trials. All number are percentage numbers without $\%$.}
\label{table4}
\resizebox{0.8\textwidth}{!}
{\begin{tabular}{crrrrrr}
 \toprule[1.1pt]
Model                  & Metric & FP32(baseline)                       & INT8          & INT4        & INT2        & INT1        \\ \hline
                       & Recall@20 (\%) & $7.14 {\color[HTML]{656565} \pm 0.04}$ & $7.12{\color[HTML]{656565} \pm 0.01}$  & $7.11{\color[HTML]{656565} \pm0.03}$ & $7.02{\color[HTML]{656565} \pm0.02}$ & $6.73{\color[HTML]{656565} \pm0.04}$ \\
\multirow{-2}{*}{KGAT} & NDCG@20 (\%)   & $8.78 {\color[HTML]{656565} \pm 0.02}$ & $8.76 {\color[HTML]{656565} \pm 0.03}$  & $8.73{\color[HTML]{656565} \pm0.02}$  & $8.63{\color[HTML]{656565} \pm0.02}$  & $8.28{\color[HTML]{656565} \pm0.03}$  \\ \hline
                       & Recall@20 (\%) & $6.83 {\color[HTML]{656565} \pm 0.02}$ & $6.81{\color[HTML]{656565} \pm 0.03}$  & $6.80{\color[HTML]{656565} \pm0.02}$ & $6.70{\color[HTML]{656565} \pm0.03}$ & $6.43{\color[HTML]{656565} \pm0.03}$ \\
\multirow{-2}{*}{KGNN} & NDCG@20 (\%)   & $7.87 {\color[HTML]{656565} \pm 0.03}$ & $7.86 {\color[HTML]{656565} \pm 0.03}$  & $7.83{\color[HTML]{656565} \pm0.02}$  & $7.74{\color[HTML]{656565} \pm0.01}$  & $7.43{\color[HTML]{656565} \pm0.02}$  \\ \hline
                       & Recall@20 (\%) & $7.79 {\color[HTML]{656565} \pm 0.03}$ & $7.78{\color[HTML]{656565} \pm 0.02}$  & $7.74{\color[HTML]{656565} \pm0.01}$ & $7.67{\color[HTML]{656565} \pm0.02}$ & $7.35{\color[HTML]{656565} \pm0.02}$  \\
\multirow{-2}{*}{KGIN} & NDCG@20 (\%)   & $9.01 {\color[HTML]{656565} \pm 0.02}$ & $8.99 {\color[HTML]{656565} \pm 0.01}$  & $8.96{\color[HTML]{656565} \pm0.02}$  & $8.84{\color[HTML]{656565} \pm0.03}$  & $8.54{\color[HTML]{656565} \pm0.03}$  \\ \toprule[1.1pt]
\end{tabular}}
\end{table*}

\subsection{Main Results}
\subsubsection{\textbf{Overall Performance.}}
The performance of our proposed TinyKG depends on the number of quantization bins $B$.  To evaluate how the quantized level affects the model performance, we vary the $B$ within $[8, 4, 2, 1]$.  Ideally, a larger $B$ can achieve the  closer behavior as full-precision settings, but with more memory footprint.  Table~\ref{table2}, Table~\ref{table3}, and Table~\ref{table4} show the performance of KGNNs on Amazon-book, MovieLens-20M, and Yelp2018, respectively.

From the tables, we can observe that our TinyKG can consistently achieve comparable performance as baselines. For example, the loss in accuracy is less than $0.3\%$ when using the INT8 quantization. And the INT2 quantization only causes a small accuracy drop ($< 2\%$) for all cases. Even with the extreme INT1 quantization, the accuracy loss is still acceptable, usually $<6\%$ in almost all experiments, in terms of both Recall@20 and NDCG@20. In contrast, adopting the INT1 or INT2 quantizations will cause a significant accuracy drop for CNNs (usually $>6\%$)~\cite{chen2021actnn}.

\begin{figure*}
	\begin{center}
	\includegraphics[width=0.8\linewidth]{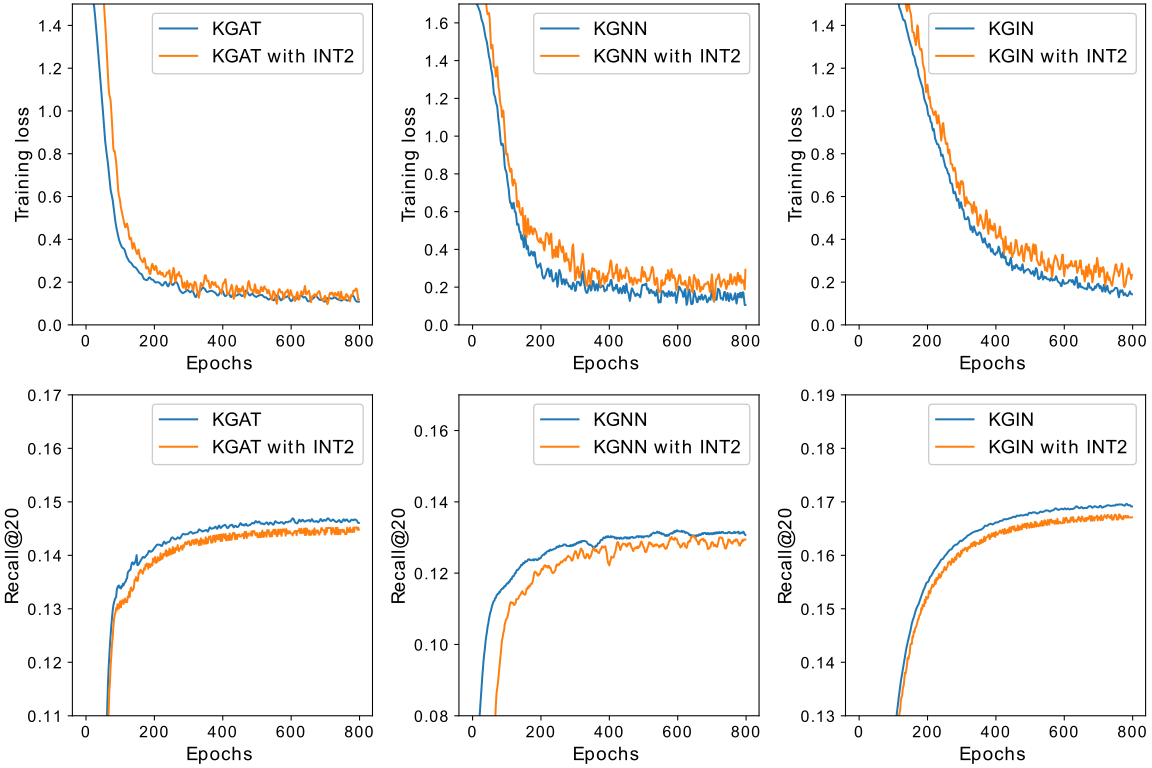}
	\end{center}
	\caption{Training curves comparison for KGNNs with/without TinyKG (INT2) for Amazon-Book.}
	\label{fig2}
\end{figure*}

In addition, we visualize the training loss curves of KGNNs with/without TinyKG in the Figure~\ref{fig2}. As can be seen, all curves of KGNNs under TinyKG with INT2 quantization are consistent with their original baselines. Therefore, the performance of proposed TinyKG can achieve quite comparable results with the state-of-the-art in terms of accuracy.


\begin{table*}[]
\caption{Comparison on the test accuracy, running time,  and memory saving on Amazon-book dataset. "Act Mem" is the memory (MB) occupied by activation maps.  "GPU Time" indicates the running time of one epoch. "Acc Loss" denotes the performance drop in term of Recall@20. All reported results are averaged over ten random trials, and the average results are reported.}
\label{mem}
\resizebox{0.86\textwidth}{!}
{\begin{tabular}{cr|rrrrr}
\toprule[1.1pt]
     &          & FP32 & INT8 & INT4 & INT2 & INT1 \\ \hline
     & Act Mem (MB) &  $1117.1$    &  $503.5(2.22\times)$    &  $377.5(2.96\times)$    &  $157.3(7.10\times)$     &   $112.3(9.95\times)$   \\
KGAT & GPU Time (Sec) &   $321.9$   &   $498.1(+54.74\%)$   &    $432.6(+34.39\%)$  &    $402.4(+25.01\%)$  &  $386.2(+19.98\%)$    \\
     & Acc Loss &    $-0\%$  &  $-0.26\%$    &  $-0.47\%$    &  $-1.26\%$    &    $-5.01\%$  \\ \hline
     & Act Mem (MB) &   $1358.6$   &   $845.3(1.61\times)$    &  $495.2(2.74\times)$     &  $187.9(7.23\times)$    &   $132.5(10.2\times)$   \\
KGNN & GPU Time (Sec)&  $303.2$    &    $467.3(+54.12\%) $  &     $413.2(+36.28\%)$ &   $379.0(+25.00\%) $   &  $337.2(+11.21\%)$    \\
     & Acc Loss &    $-0\%$  &  $-0.21\%$    &  $-0.41\%$    &  $-1.31\%$    &    $-5.30\%$  \\ \hline
     & Act Mem (MB) &   $1274.3$   &     $674.0(1.89\times)$ &  $457.6(2.78\times)$    &    $168.2(7.58\times)$  & $119.2(10.6\times)$     \\
KGIN & GPU Time (Sec) &  $378.8$    &   $524.2(+38.38\%) $   &     $471.3(+24.42\%)$ &  $445.0(+17.72\%)$    &     $410.4(+8.40\%)$ \\
       & Acc Loss &    $-0\%$  &  $-0.23\%$    &  $-0.44\%$    &  $-1.43\%$    &    $-5.42\%$  \\ \toprule[1.1pt]
\end{tabular}}
\end{table*}

	\begin{figure*}
		\begin{center}
		\includegraphics[width=0.80\linewidth]{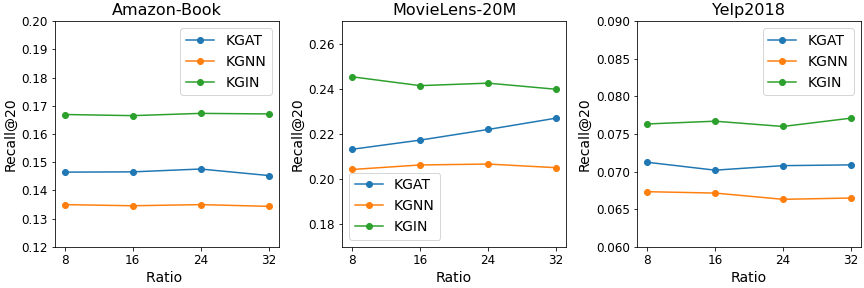}
		\end{center}
		\caption{ The sensitivity study of TinyKG (INT2) to the ratio $\frac{d}{B^2}$ on three datasets.}
			\label{fig3}
	\end{figure*}

\begin{table*}[t]
\caption{The performance (Recall@20) of KGNNs trained on the Amazon-book dataset under
stochastic rounding (SR) and nearest rounding (NR). All results are averaged over ten random trials. All number are percentage numbers without $\%$, and '-' means  the algorithm fails to converge with $>40\%$ performance drop}.
\label{table20}
\resizebox{0.8\textwidth}{!}
{\begin{tabular}{crrrrrr}
 \toprule[1.1pt]
Model                  & Metric & FP32(baseline)                       & INT8          & INT4        & INT2        & INT1        \\ \hline
                       & SR(\%) & $14.85  {\color[HTML]{656565} \pm 0.15}$ & $14.81{\color[HTML]{656565} \pm0.12}$   & $14.78{\color[HTML]{656565} \pm0.14}$ & $14.66{\color[HTML]{656565} \pm0.14}$ & $14.08{\color[HTML]{656565} \pm0.15}$ \\
\multirow{-2}{*}{KGAT} & NR(\%)   &  $14.85  {\color[HTML]{656565} \pm 0.15}$                       & $10.04 {\color[HTML]{656565} \pm0.12}$   & -  & -  & -  \\ \hline
                       & SR(\%) & $13.64 {\color[HTML]{656565} \pm 0.16}$                        & $13.61 {\color[HTML]{656565} \pm 0.18}$ & $13.58{\color[HTML]{656565} \pm0.12}$ & $13.46{\color[HTML]{656565} \pm0.12}$ & $12.91{\color[HTML]{656565} \pm0.13}$ \\
\multirow{-2}{*}{KGNN} & NR(\%)   &  $13.64 {\color[HTML]{656565} \pm 0.16}$                       & $9.87{\color[HTML]{656565} \pm 0.17}$  & - & - & - \\ \hline
                       & SR(\%) & $16.89{\color[HTML]{656565} \pm0.16}$                          & $16.85 {\color[HTML]{656565} \pm 0.13}$ & $16.82{\color[HTML]{656565} \pm0.12}$ & $16.65{\color[HTML]{656565} \pm0.12}$ & $15.97{\color[HTML]{656565} \pm0.11}$  \\
\multirow{-2}{*}{KGIN} & NR(\%)   &$16.89{\color[HTML]{656565} \pm0.16}$    &    $13.89{\color[HTML]{656565} \pm0.24}$                     & - & - & - \\ \toprule[1.1pt]
\end{tabular}}
\end{table*}

\subsubsection{\textbf{Memory and Speed.}}
  As discussed before, our TinyKG is able to reduce the training memory footprint. However, we notice that TinKG slightly slows down the training speed. The extra time cost comes directly from the (de)quantization. Table~\ref{mem} shows the trade-off among the memory, speed, and performance. Here we only show the results on Amazon-Book, and the observations are similar for other two datasets. We simply omit their results.
  
  From Table~\ref{mem},  we note that lower precision quantization can save more memory footprint, with smaller time overhead, but with larger performance drop. INT1 quantization can aggressively reduces the memory footprint with $7.1 \times$, yet with around $5\%$ performance drop in many situations. In practice, one can safely adopt INT2 quantization for training KGNNs to achieve a good balance. For instance, KGAT with INT2 quantization generally reduces the memory footprint of activation maps with $7.1 \times$, but merely with $1.26\%$ loss in accuracy, while the time overhead is roughly $25\%$; KGIN with INT2 quantization reduces $7.58 \times$ the memory footprint of activations with just $1.43\%$ performance drop, and the time overhead roughly ranges from $17\%$to $25\%$.

As such, one can choose a larger batch size or a more complex network architecture to fully utilize the power of neural message-passing mechanisms with the relaxed memory footprint restriction~\cite{l.2018dont}. That is, our TinyKG unveils the great value by providing more choices to explore the design of KGNNs.

\subsubsection{\textbf{The effects of ratio $\frac{d}{B^2}$.}}

From the property of our quantized strategy given in Proposition 1, the calculated gradients are unbiased. However, the quantization also imposes extra variance to the calculated gradients, where the variance linearly scales with the ratio $\frac{d}{B^2}$. Note that the $R^{(l)}_v$ is tuned on-the-fly. To quantitatively study the effect of the extra variance, we fix $B=2$ (INT2 quantization), and vary $d$ within $\{32,64,96,128\}$. We show the performance of KGNNs on three datasets. 

From Figure~\ref{fig3}, we observe that the  loss  in  accuracy is generally negligible. This can be explained by the following: 1) the model depth of KGNNs is much smaller than CNNs due to the over-smoothing problem. The accumulate quantized errors are thus relatively small according to Eq. (\ref{eq6}). 2) Although the quantized steps introduce extra noise in gradients during the backward pass, recent work~\cite{godwin2022simple} finds that simple noisy regularization can be an effective way to address over-smoothing for message-passing schema, while CNNs do not have such property~\cite{chen2021actnn}. Therefore, KGNNs are much more noise-tolerant, allowing for lower bit quantizations.

\subsubsection{\textbf{Stochastic Rounding vs. Nearest Rounding.}}

To explore the effect of stochastic rounding in TinyKG, we compare it with the commonly used nearest rounding to evaluate how different rounding algorithms affect the performance. Both stochastic rounding and nearest rounding share similar memory footprint during training since the memory cost is  determined by the number of quantization bins. In this work, we mainly compare their performance on Amazon-Book in terms of Recall@20. As Table~\ref{table20} shows, nearest rounding fails to converge at most cases. This suggests that stochastic rounding is important to guarantee good performance in TinyKG.

\section{Conclusion and Future Work}

In this paper, we propose TinyKG, a simple-yet-effective framework for training KGNNs with compressed activation maps. Specifically, we leverage a uniform quantization with a stochastic rounding algorithm to efficiently compress the intermediate activations while training KGNNs. In addition, we verify the unbiasedness and low variance of our introduced quantization. To evaluate the performance of our method, we conduct comprehensive experiments over three real-world datasets for downstream recommendation tasks. In the experiments, we systematically analyze the trade-off of our strategy among the memory-saving, time overhead, and accuracy drop. The experimental results demonstrate that TinyKG can extensively reduce the GPU's memory footprint, while merely incurring a slight time overhead and performance drop. In addition, we show that our proposed TinyKG can be successfully applied to  existing KGNNs without much extra engineering. 

As our future work, we would like to design self-supervised KGNNs under our TinyKG. The  graph contrastive learning framework~\cite{you2020graph} can  learn  local and global node representations to better capture structure information. Nevertheless, the contrastive frameworks usually require a large batch size of  comparing pairs to obtain more accurate estimation of the contrastive loss, leading to large GPU memory. We plan to reduce the training memory footprint of constrative framework using our compressed technique. Moreover, our TinyKG is orthogonal to most of existing  techniques, including distributed training~\cite{ren2021smore}, compression~\cite{han2015deep_compression} and gradient compression~\cite{lin2018dgc}. We are interested to explore the potential benefits of integrating TinyKG with those efforts.

\section*{Appendix}
\label{sec:append}

The  quantization of Eq. (\ref{deq}) and Eq. (\ref{deq2}) for activation $\mathbf{e}_v^{(l)} \in \mathbb{R}^d$ is unbiased, its expectation and variance are:
\begin{equation*}
\mathbb{E}[\hat{\mathbf{e}}^{(l)}_v]=\mathbb{E}[\textup{Dequant}(\textup{Quant}(\mathbf{e}^{(l)}_v))] = \mathbf{e}^{(l)}_v, \quad \textup{Var}[\hat{\mathbf{e}}^{(l)}_v] \le \frac{d[R^{(l)}_v]^2}{4B^2}.
\end{equation*}
\begin{proof}
Based on the definition of  stochastic rounding~\cite{gupta2015deep,courbariaux2015binaryconnect} and the fact that $\lceil x \rceil = 1 +  \lfloor x \rfloor$, we have:
\begin{equation*}
\begin{aligned}
\mathbb{E}[ \lfloor x\rceil] &=  \lceil x \rceil \cdot  (x -\lfloor x \rfloor) +  \lfloor x \rfloor \cdot (1- (x -\lfloor x \rfloor)) \\
& = (1 + \lfloor x \rfloor)\cdot  (x -\lfloor x \rfloor) +  \lfloor x \rfloor \cdot (1- x  + \lfloor x \rfloor) \\
& = x
\end{aligned}
\end{equation*}
Similar, we have:
\begin{equation*}
\begin{aligned}
\mathbb{E}[\hat{\mathbf{e}}^{(l)}_v]&=\mathbb{E}[\textup{Dequant}(\textup{Quant}(\mathbf{e}^{(l)}_v))] \\
&= \mathbb{E}\left[ \frac{R_v^{(l)}}{B} \cdot \left\lfloor \frac{\mathbf{e}^{(l)}_v - Z^{(l)}_v}{R^{(l)}_v}B \right\rceil + Z_v^{(l)} \right] \\
& = \frac{R_v^{(l)}}{B} \cdot \mathbb{E}\left[  \left\lfloor \frac{\mathbf{e}^{(l)}_v - Z^{(l)}_v}{R^{(l)}_v}B \right\rceil  \right] +  Z_v^{(l)}\\
&= \frac{R_v^{(l)}}{B} \cdot \left( \frac{\mathbf{e}^{(l)}_v - Z^{(l)}_v}{R^{(l)}_v}B \right)   +  Z_v^{(l)} \\
& = \mathbf{e}^{(l)}_v.
\end{aligned}
\end{equation*}

For variance,
\begin{equation*}
\begin{aligned}
\text{Var}[ \lfloor x\rceil] &=  (\lceil x \rceil - x)^2 \cdot  (x -\lfloor x \rfloor) +  (\lfloor x \rfloor-x)^2 \cdot (1- x +\lfloor x \rfloor) \\
& = (1 -(x- \lfloor x \rfloor ))^2 \cdot  (x -\lfloor x \rfloor) +  (x-\lfloor x \rfloor)^2 \cdot (1- (x -\lfloor x \rfloor)) \\
& =  -(x- \lfloor x \rfloor )^2 + (x- \lfloor x \rfloor )
\end{aligned}
\end{equation*}

As such, let $\bar{\mathbf{e}}=[\bar{e}_1, \cdots, \bar{e}_d] = \frac{\mathbf{e}^{(l)}_v - Z^{(l)}_v}{R^{(l)}_v}B$, we have:
\begin{equation*}
\begin{aligned}
\textup{Var}[\hat{\mathbf{e}}^{(l)}_v] =& \text{Var}\left[ \frac{R_v^{(l)}}{B} \cdot \left\lfloor \frac{\mathbf{e}^{(l)}_v - Z^{(l)}_v}{R^{(l)}_v}B \right\rceil + Z_v^{(l)} \right] \\
& =\frac{[R_v^{(l)}]^2}{B^2} \cdot \text{Var}\left[  \cdot \left\lfloor \frac{\mathbf{e}^{(l)}_v - Z^{(l)}_v}{R^{(l)}_v}B \right\rceil \right] 
\\
&= \frac{[R_v^{(l)}]^2}{B^2} \cdot \sum_{i=1}^{d} \text{Var}[  \lfloor  \bar{e}_i  \rceil ] \\
&=  \frac{[R_v^{(l)}]^2}{B^2} \cdot \sum_{i=1}^{d}[  -(\bar{e}_i- \lfloor \bar{e}_i \rfloor )^2 + (\bar{e}_i- \lfloor \bar{e}_i \rfloor ) ]\\
&= \frac{[R_v^{(l)}]^2}{B^2} \cdot \sum_{i=1}^{d}[  -(\bar{e}_i- \lfloor \bar{e}_i \rfloor -\frac{1}{2})^2 + \frac{1}{4} ]\\
&\le \frac{d[R^{(l)}_v]^2}{4B^2}.
\end{aligned}
\end{equation*}
The inequality always holds since $\bar{e}_i- \lfloor \bar{e}_i \rfloor\in [0, 1]$. And the upper bound is achieved when  $\bar{e}_i- \lfloor \bar{e}_i \rfloor = \frac{1}{2}$.

\end{proof}

\bibliographystyle{ACM-Reference-Format}
 
\bibliography{my}

\end{document}